\documentclass[proceedings, preprint]{rmaa}

\usepackage{paralist}

\usepackage{psfrag,color}

\SetYear{2022}
\SetConfTitle{Prospects for low-frequency radio astronomy in S. A.}

\title{The Multipurpose Interferometric Array and the development of its technological demonstrator} 

\author{
   G. Gancio\altaffilmark{1}, 
   G.~E. Romero\altaffilmark{1}, 
   P. Benaglia\altaffilmark{1}, 
   J.~M. González\altaffilmark{1}, 
   E. Rasztocky\altaffilmark{1}, 
   H. Command\altaffilmark{1}, 
   G. Valdez\altaffilmark{1}, 
   E. Tarcetti\altaffilmark{1}, 
   F. Hauscarriaga\altaffilmark{1}, 
   P. Alarcón\altaffilmark{1}, 
   F. Aquino\altaffilmark{1}, 
   M. Alí\altaffilmark{1}, 
   D. Capuccio\altaffilmark{1}, 
   L.~F. Cabral\altaffilmark{1}, 
   M. Contreras\altaffilmark{1}, 
   E. Díaz\altaffilmark{1}, 
   N. Duarte\altaffilmark{1},
   L.~M. García\altaffilmark{1}, 
   D. Perilli\altaffilmark{1}, 
   P. Otonello\altaffilmark{1}, 
   S. Spagnolo\altaffilmark{1} 
    }

\altaffiltext{1}{Instituto Argentino de Radioastronomía (CONICET -- CIC -- UNLP), C.C.5, (1984) Villa Elisa, Buenos Aires, Argentina (ggancio@iar.unlp.gov.ar).}

\shortauthor{Gancio 
et al.}
\shorttitle{Multipurpose Interferometric Array (MIA)}

\listofauthors{G. E. Romero, G. Gancio, P. Benaglia, J. M. González,
  E. Rasztocky, H. Command, G. Valdez, E. Tarcetti, F. Hauscarriaga, P. Alarcón, F. Aquino, M. Alí, L. F. Cabral, D. Capuccio, M. Contreras, E. Díaz, N. Duarte, L. M. García, D. Perilli, P. Otonello, S. Spagnolo}
\indexauthor{Romero, G. E.}
\indexauthor{Gancio, G}
\indexauthor{Benaglia, P.} 
\indexauthor{González, J. M.}
\indexauthor{Rasztocky, E.} 
\indexauthor{Command, H.} 
\indexauthor{Valdez, G.}
\indexauthor{Tarcetti, E.} 
\indexauthor{Hauscarriaga, F.} 
\indexauthor{Alarcón, P.} 
\indexauthor{Aquino, F.}
\indexauthor{Alí, M.}
\indexauthor{Capuccio, D.}
\indexauthor{Contreras, M.}
\indexauthor{Díaz, E.}
\indexauthor{Duarte, N.}
\indexauthor{García, L.~M.}
\indexauthor{Cabral, L.~F.}
\indexauthor{Perilli, D.}
\indexauthor{Otonello, P.}
\indexauthor{Spagnolo, S.}

\abstract{We present a proposal for the construction and development of a new instrument for radio astronomical observations based on interferometric techniques, that will provide high angular resolution in the 21 cm band, with the intention of improving and extending the current performance of the instruments used at the Argentine Institute of Radio Astronomy. This will allow internationally competitive scientific research and the acquisition of cutting-edge scientific and technological know-how in the aforementioned techniques, enabling interferometric measurements and the development of very long baseline or VLBI techniques. This project is called MIA, an acronym for ``Multipurpose Interferometric Array''.}

\resumen{Presentamos una propuesta para la construcción y desarrollo de un nuevo instrumento para observaciones radioastronómicas basado en técnicas interferométricas, que proporcionará alta resolución angular en la banda de 21 cm, con la intención de mejorar y ampliar las prestaciones actuales de los instrumentos utilizados en el Instituto Argentino de Radioastronomía. Esto permitirá realizar investigación científica competitiva a nivel internacional y adquirir conocimientos científicos y tecnológicos de vanguardia en las técnicas mencionadas, posibilitando mediciones interferométricas y el desarrollo de técnicas de línea de base muy larga o VLBI. El proyecto se denomina MIA por sus siglas en inglés ``Multipurpose Interferometric Array''.}

\addkeyword{techniques: interferometric}
\addkeyword{instrumentation: interferometers}
\addkeyword{radio continuum: general}
\addkeyword{pulsars: general}

\begin{document}
\maketitle

\section{Multipurpose Interferometric Array, the big picture}
\label{sec:intro}
The Multipurpose Interferometric Array, or MIA, is a novel radio astronomical observing instrument based on interferometric techniques that will provide high angular resolution in the 21 cm band, and it is the first of its kind in South America. The project aims to improve and expand the scientific research (e.g. \citealp{carignan2013}, \citealp{colomb1980}, \citealp{testori2001}) and instruments used at the Argentine Institute of Radio Astronomy 
(\citealp{gancio2020}), allowing researchers to carry out cutting-edge scientific research at an international level and to acquire practical scientific and technological know-how in interferometric techniques in Argentina.

MIA is designed to be a low-cost instrument in its initial phase, with the ability to expand and upgrade over time. The instrument will consist of several radio antennas connected to a central computer that will process the data and generate the images. The final configuration may include up to 64 antennas arranged in a configuration such to achieve important angular resolution.

MIA will have very high sensitivity, allowing researchers to detect faint radio sources with high precision. The instrument will be capable of observing at frequencies between 1 GHz and 2.3 GHz with the capability of add as a sub array low frequency antennas, enabling a wide range of scientific investigations.

The scientific potential of MIA will cover a wide range of research areas in astrophysics. Among the research areas for MIA's capabilities are the detection of transient counterparts to gamma-ray bursts, the study of fast radio bursts, the timing of pulsars, and the observation of flares in magnetars. In addition, MIA will allow the spectro-temporal study of X-ray binaries and microquasars, and the identification of the counterpart of unidentified gamma-ray sources.

Other research areas where MIA can make significant contributions include multi-frequency variability studies of active galactic nuclei, the study of the morphology and spectral distribution of supernova remnants, and the mapping of continuous non-thermal extended sources. MIA's sensitivity will also allow the study of the HI line at cosmological distances to characterize the reionization epoch of the Universe, and the study of the interstellar medium with high angular resolution.

MIA will also allow researchers to study the physical and kinematic characterization of ionized hydrogen regions, OH maser variability studies in star forming regions, and the study of extended star forming regions. This will lead to new insights into the dynamics and physical conditions of these astrophysical phenomena.

MIA's capabilities and sensitivity make it an ideal instrument for collaborative projects with other international observatories, allowing researchers to conduct comprehensive studies of various astrophysical phenomena. In addition, MIA's location in South America will provide the scientific community with unique observing opportunities. MIA will play an important role in the expansion of astrophysical research in Argentina and contribute to the global scientific community's efforts to deepen our understanding of the Universe.

\begin{figure}[!t]
  \includegraphics[width=\columnwidth]{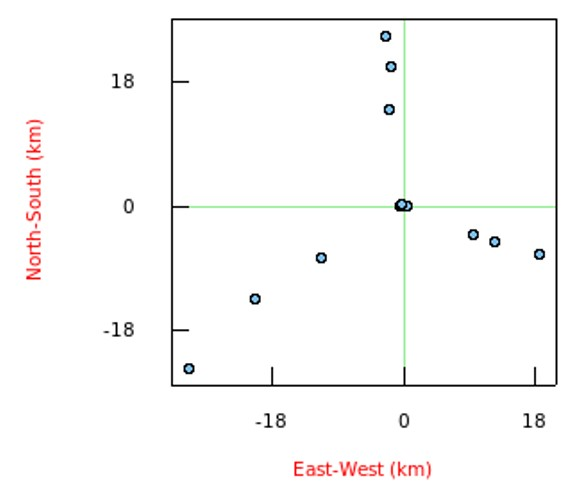}
  \caption{View of 16 antennas in a star-shaped proposal.}
  \label{fig:16-ant}
\end{figure}

\begin{figure}[!t]
  \includegraphics[width=\columnwidth]{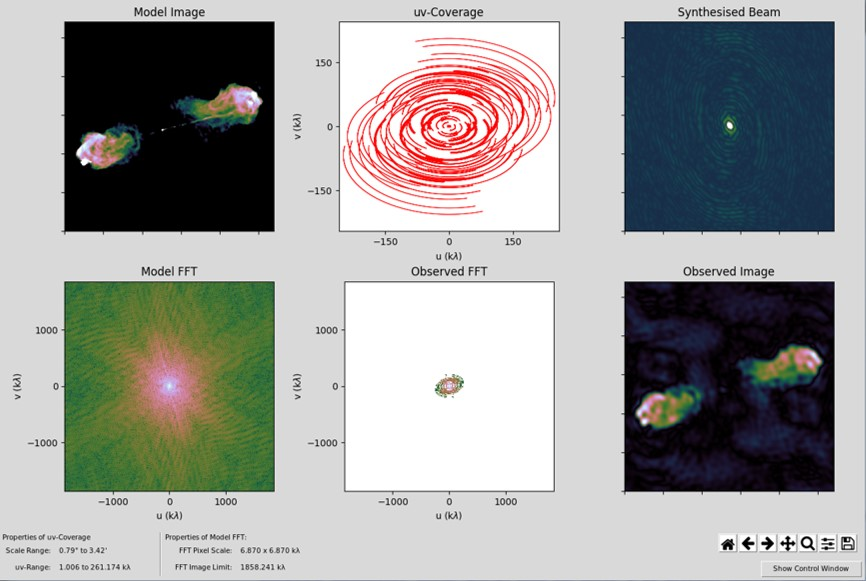}
  \caption{Response of 16 antennas in the layout proposed in Figure~\ref{fig:16-ant}.}
  \label{fig:sim16}
\end{figure}

Figure~\ref{fig:16-ant} shows one of the possible configurations under study for the location of the Phase-1 sixteen antennas; with this configuration, using a simulation software called The Friendly Virtual Radio Interferometer\footnote{https://crpurcell.github.io/friendlyVRI/}, we can evaluate the response to an observation shown in Figure~\ref{fig:sim16}.

\section{Road to Construction and Financial Support}
\label{sec:construction}

The construction and financial support of the MIA project have been planned to ensure its success as a low-cost instrument. Compared to similar instruments around the world, MIA is designed to be manufactured and assembled locally, providing an open opportunity for local companies to present themselves as technology providers. This approach will also contribute to the development of the local economy and the creation of new employment opportunities.

One of the objectives of the MIA project is to obtain financial support through collaboration with international partners. This financial support will be used to fund the various phases of the project, including the construction and installation of the antennas, the central node, and the data processing and analysis systems. The participation of international partners will also ensure the success and sustainability of the project.

In addition to financial support, the MIA project offers an open opportunity for local institutions to become partners, especially during the site selection phase. The involvement of local institutions in this phase of the project will ensure that the selected site meets the necessary requirements for the optimal functioning of the interferometer. This will include considerations such as accessibility, existing infrastructure, radio interference, and others.

\begin{figure}[!t]
  \includegraphics[width=\columnwidth]{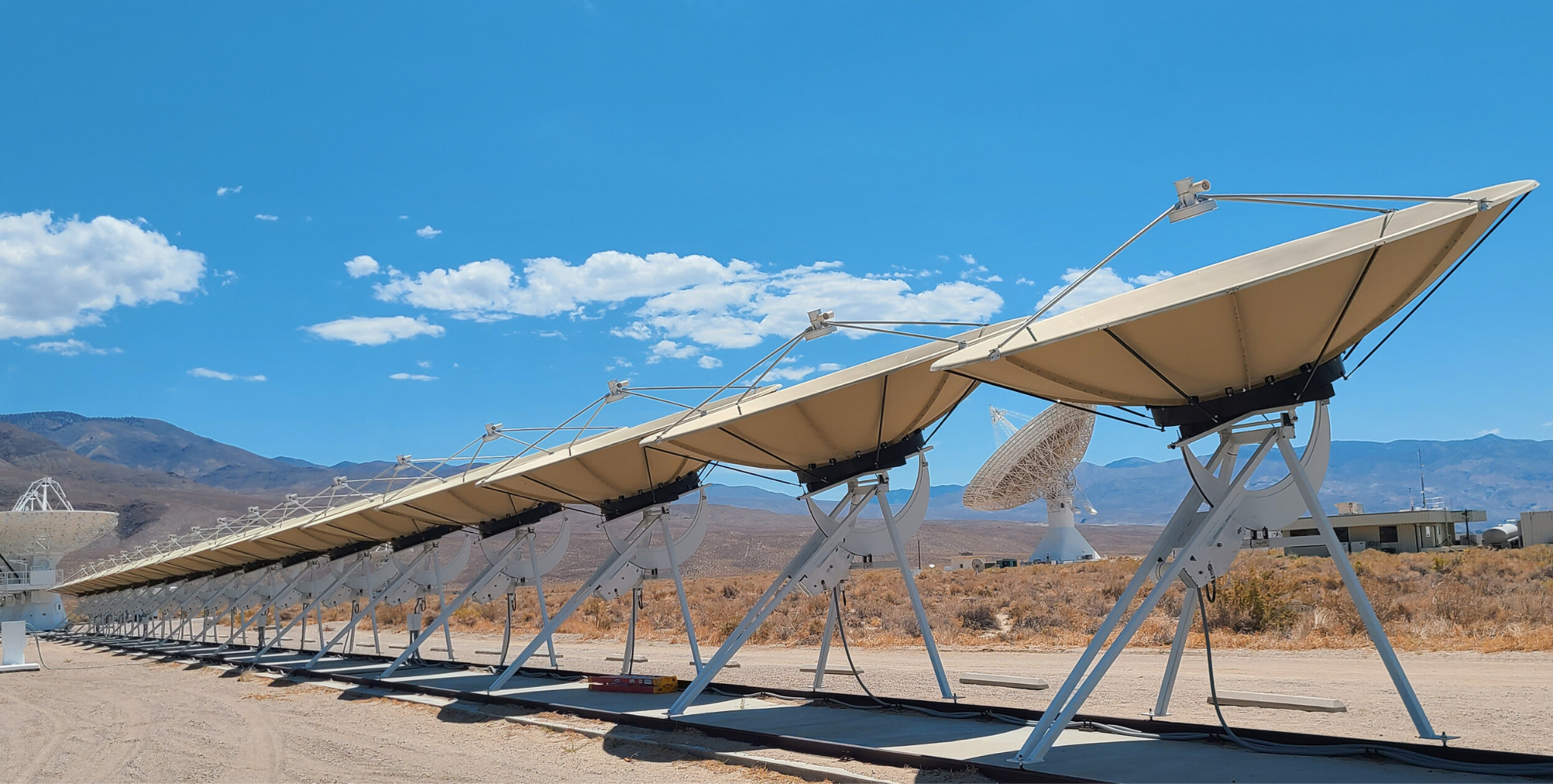}
  \caption{DSA-110 antenna locations.}
  \label{fig:dsa110}
\end{figure}

To ensure the success of the MIA project, we have learned from previous projects such as \citealp{deboer2017}, \citealp{hickish2019} and \citealp{kocz2019}, and from instruments with similar structures such as the Deep Synoptic Array 110 (DSA-110)\footnote{https://www.deepsynoptic.org/instrument} (Fig.~\ref{fig:dsa110}) and the Karoo Array Telescope (KAT-7)\footnote{http://public.ska.ac.za/kat-7}. This learning has helped us to understand and adapt systems specifically for MIA while also taking into account the unique characteristics and requirements of the project.

The technical summary of the MIA project includes a total of 16 antennas for Phase 1, each with a diameter of 5 m and an alt-azimuth mount. The minimum baseline for the antennas is approximately 50 meters, while the maximum baseline is 55 km. The interferometer has a minimum angular resolution of 1.5 arcsec at 1420 MHz / 50 km and a receiver temperature of less than 50 K.

The MIA project operates in the frequency range of 1 to 2.3 GHz and digitizes the signals received in each antenna. The final bandwidth of the system is 1000 MHz and the correlation of the signals takes place in a central node. In addition, the interferometer has the capability to add antennas to both the central core and the external antennas, allowing for future expansion and development of the system.

The full technical summary is presented below:
\begin{asparaitem}
\item Number of antennas for Phase 1: 16
\item Diameter of each antenna = 5 m (with alt-azimuth mount)
\item Minimum baseline = ~50 m
\item Maximum baseline = 55 km
\item Minimum angular resolution = 1.5 arcsec at 1420 MHz / 50 km
\item Receiver temperature = 5~K
\item Operating frequency = 1 to 2.3 GHz
\item Digitization in each antenna
\item Final bandwidth of 1000 MHz
\item Correlation in a central node
\item Ability to add antennas both to the central node and beyond the external antennas.
\end{asparaitem}

\section{Short term plan}
\label{sec:plan}

\begin{figure*}[!t]
  \centering
  \includegraphics[width=1\textwidth]{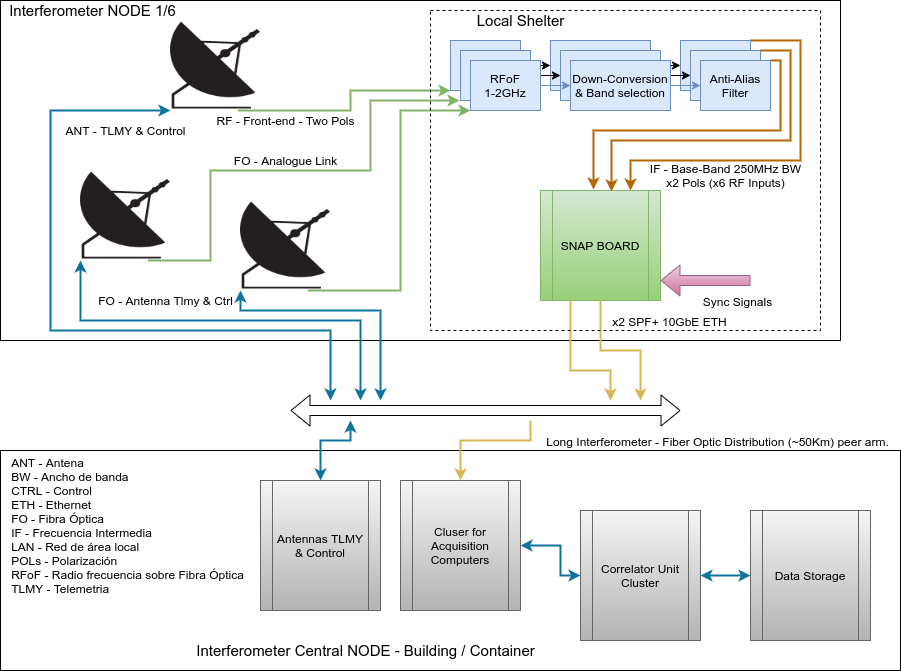}
  \caption{Pathfinder block diagram.}
  \label{fig:mia3block}
\end{figure*}

The short-term plan for the MIA project, which is currently underway, is to design and build a pathfinder instrument consisting of three antennas to be installed in the IAR, with multiple objectives as a technology demonstrator, human resource training, and scientific data validation. The primary goal is to develop the technical skills necessary to build a larger interferometer, which will include the development and testing of interferometric observing techniques and the creation of data acquisition and processing programs.
Figure~\ref{fig:mia3block} shows a block diagram of all the systems involved in the Pathfinder design.

\begin{figure*}[!t]
  \centering
  \includegraphics[width=\textwidth]{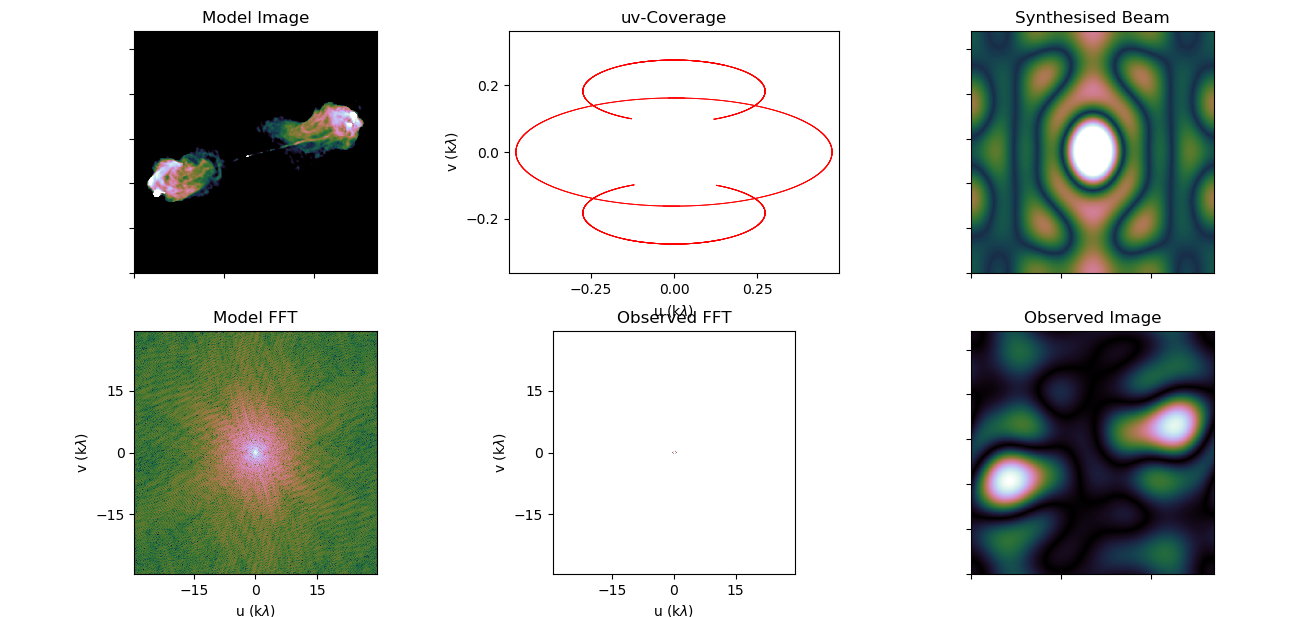}
  \caption{Simulated response of the pathfinder.}
  \label{fig:mia3sim}
\end{figure*}

\begin{figure}[!t]
  \includegraphics[width=\columnwidth]{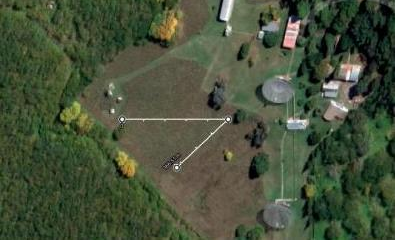}
  \caption{Proposed location for Pathfinder antennas within the IAR.}
  \label{fig:mia3site}
\end{figure}

 For the Pathfinder we did the same kind of simulation to understand their response using a 100~m baseline; Fig.~\ref{fig:mia3sim} shows the results of the simulation while  Fig.~\ref{fig:mia3site} shows the proposed location of the elements to be installed at IAR.

Another goal of the short-term plan is to train new technical and scientific resources through calls to the university community for professional internships, graduate theses, and Ph.D.s in engineering and astronomy. This training will help ensure the success and sustainability of the project by fostering the development of skilled professionals who can contribute to the future growth and expansion of the MIA project.

The scientific objective of the short-term plan is to validate the scientific data collected by the MIA interferometer. This objective is non-competitive and focuses on the ability of the scientific community to verify and reproduce the data collected by the interferometer.

Funding for the Pathfinder project is provided by CONICET through one so-called ``PUE'' program (Proyecto de investigación de Unidad Ejecutora) for research centers. The majority of the funds will be used for the development of the complete system, which includes the construction and installation of the antennas, front-end, back-end, digitizer and processing systems.

The MIA Pathfinder organization and key requirements include the design and construction of the entire antenna structure at the IAR. The parabolic dish diameter is 5~m, and the mount type and movements are Alt (0$^\circ$ to 90$^\circ$) - Azimuth (0$^\circ$ -- 350$^\circ$). The front-end and back-end systems are designed and built at the IAR, with linear polarization H plus V and an operating frequency range of 1 GHz to 2 GHz. The receiver temperature is less than 65~K and the radio frequency (RF) link is analog fiber optic (RFoF).

\begin{figure}[!t]
  \includegraphics[width=0.95\columnwidth]{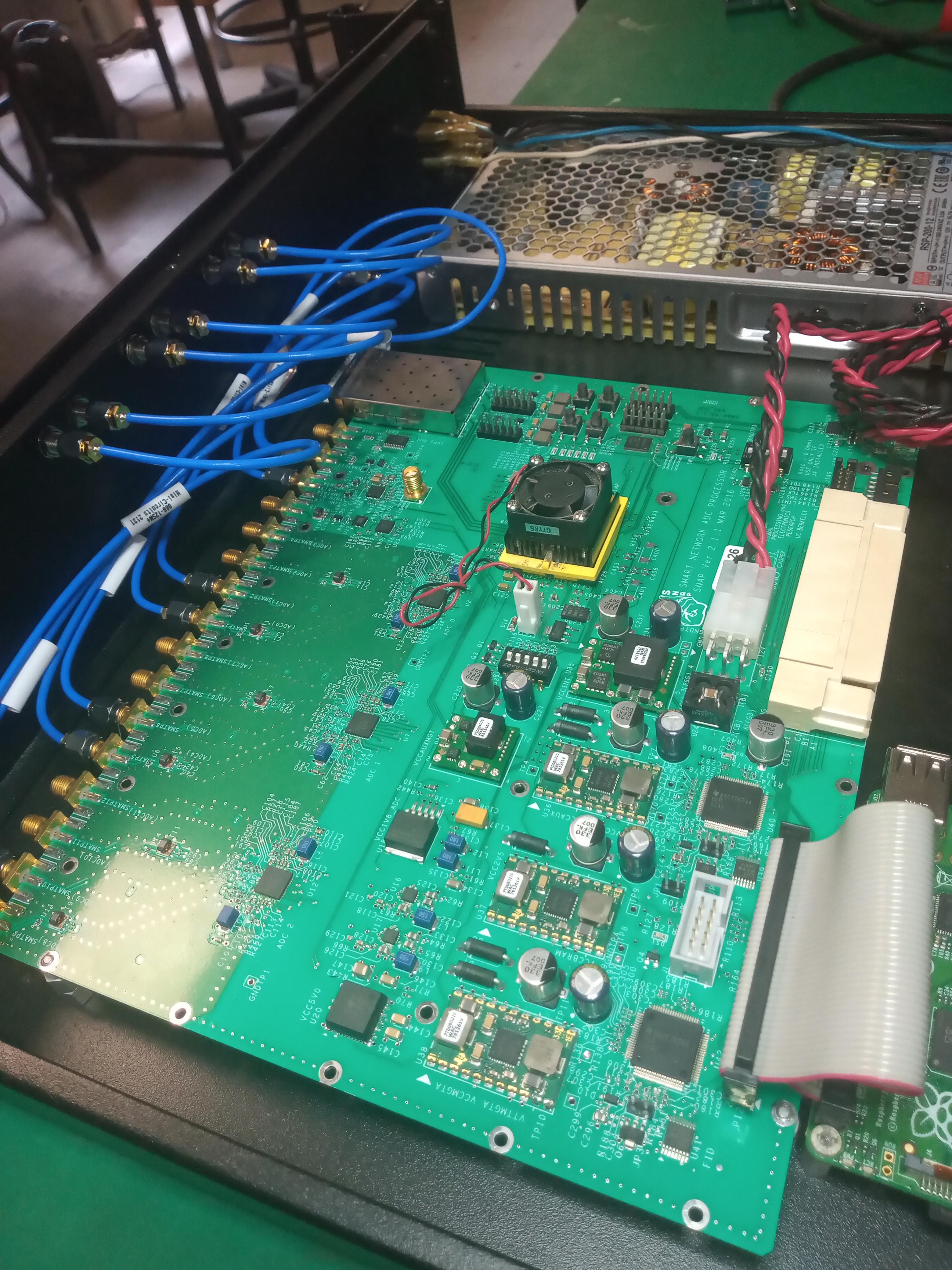}
  \caption{SNAP board for Pathfinder digitizer.}
  \label{fig:snap}
\end{figure}

The digitizer and processing systems use the CASPER\footnote{https://casper.berkeley.edu/} toolflow and are based on the SNAP board (Fig.~\ref{fig:snap}) with a bandwidth of 250 MHz x 3 antennas x 2 polarizations, resulting in a total bandwidth of 1500 MHz. 
CPU/GPU are used  
for processing and data reduction; the data is transmitted over a 10~Gbps network.

 The personnel involved in the MIA Pathfinder project include a Ph.D. student for antenna distribution and optimization, IAR technical staff for electronics, mechanics, IT, and Electronic Engineering students from the La Plata National University, who are doing their thesis on related topics such as RF systems, digitizers, and correlation, and a couple of scientific researchers.

\section{Current Status}
\label{sec:status}
\begin{figure*}[!t]
  \centering
  \includegraphics[width=\textwidth]{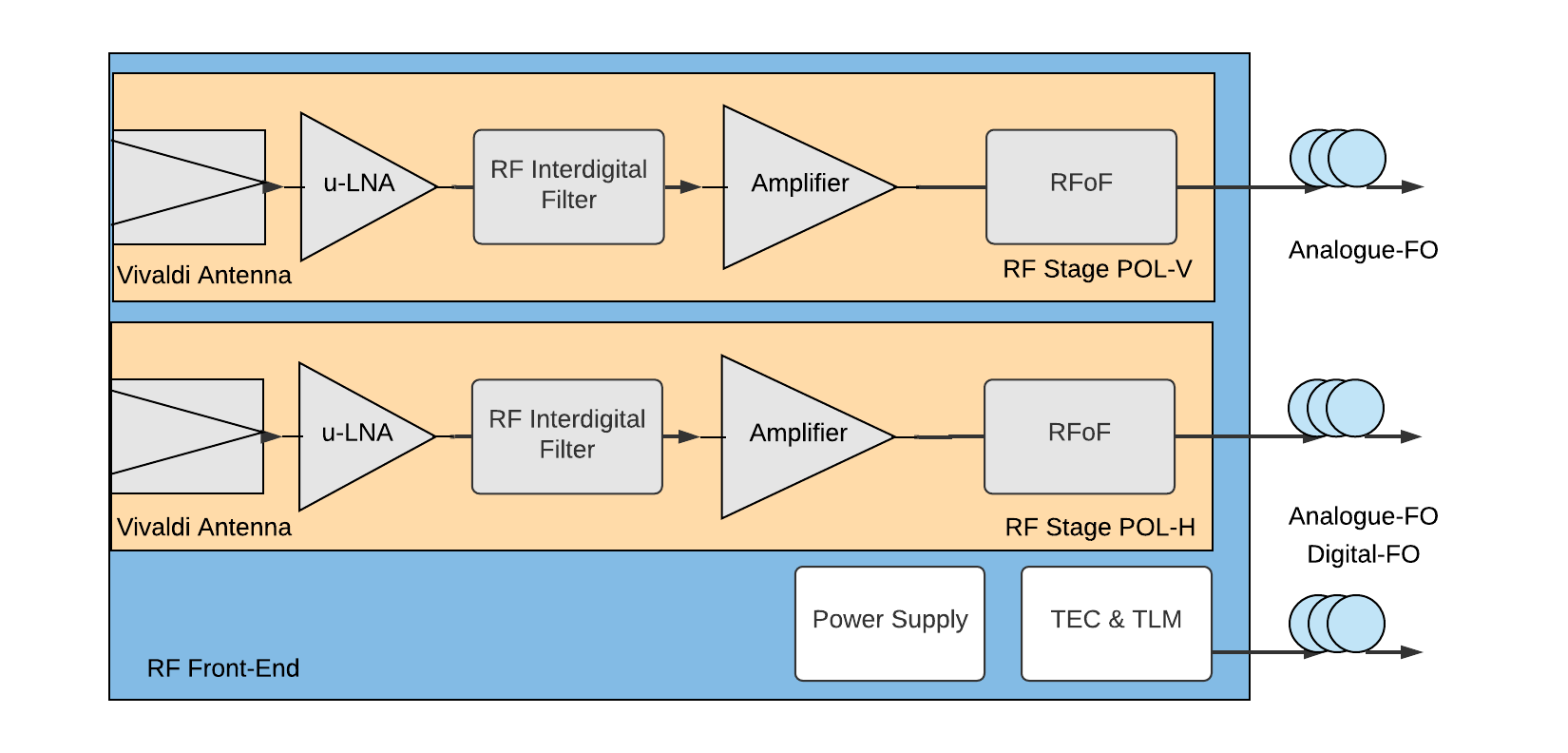}
  \caption{RF front-end block diagram.}
  \label{fig:fend}
\end{figure*}

The MIA project has made significant progress up to its current state, and included several critical components in the design, test, and integration process. Especially, in the design of its radio frequency front-end, which is currently being tested for functionality and integration with the telemetry electronics. This front-end design is a critical component of the MIA-Phase 1 signal processing chain as it receives incoming signals from the 16 antennas and converts them to digital signals for processing. A basic block diagram of the radio frequency front-end is shown in Fig.~\ref{fig:fend}. The antenna feed is designed as a board antenna called Vivaldi, which has two linear polarizations. The RF stages consist of low noise amplifiers and bandwidth limiting filters and the RFoF modules, all designed, integrated and tested at the IAR.

Figure~\ref{fig:antcai} shows the Vivaldi antenna being tested in the IAR anechoic chamber.

\begin{figure}[!t]
  \includegraphics[width=\columnwidth]{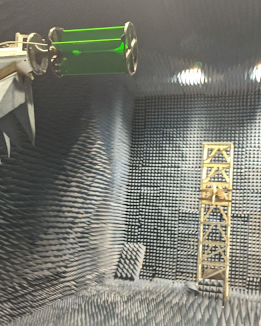}
  \caption{Measurement of the Vivaldi antenna at IAR anechoic chamber.}
  \label{fig:antcai}
\end{figure}

\begin{figure}[!t]
  \includegraphics[width=0.95\columnwidth]{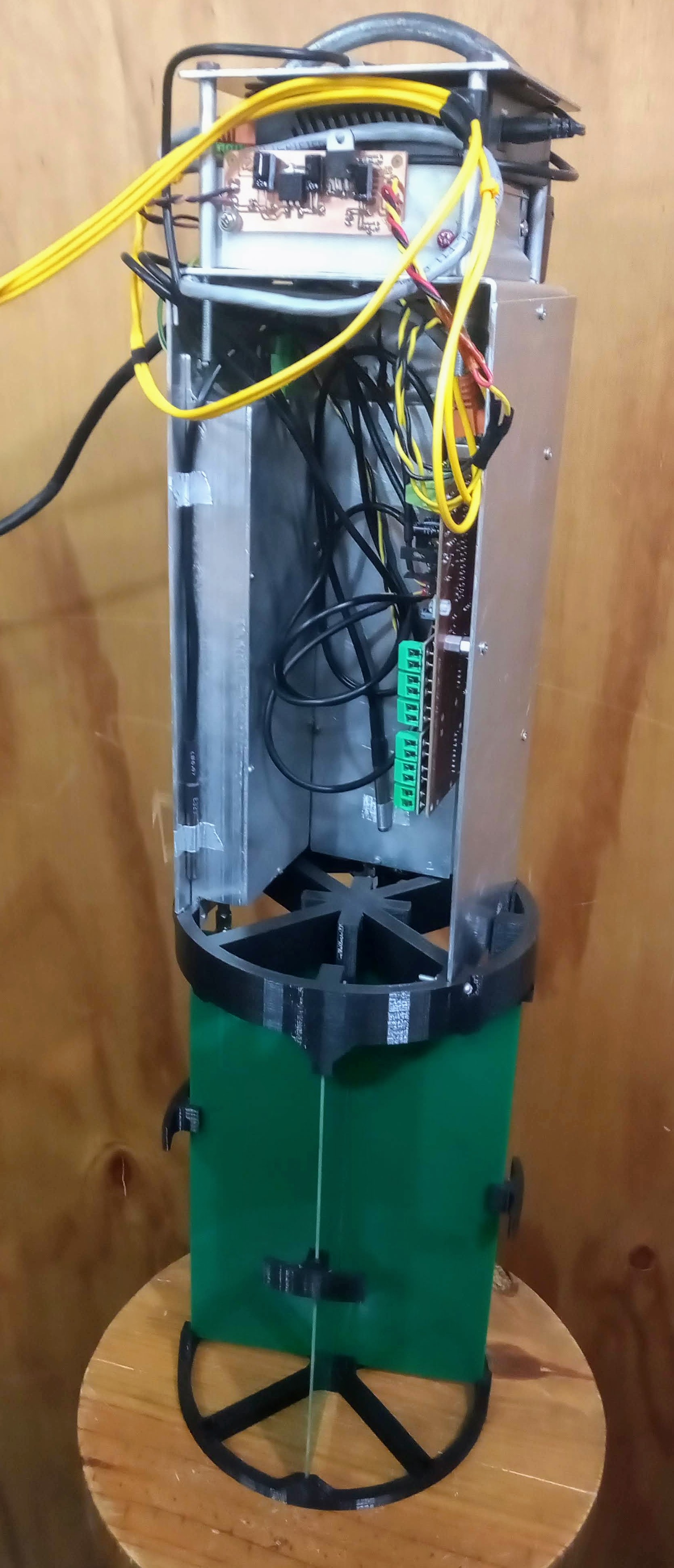}
  \caption{First front-end assembly.}
  \label{fig:antcai}
\end{figure}

\begin{figure}[!t]
  \includegraphics[width=\columnwidth]{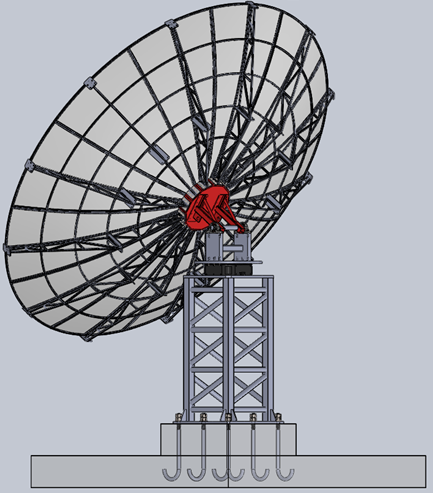}
  \caption{Antenna CAD design}
  \label{fig:antcad}
\end{figure}

\begin{figure}[!t]
  \includegraphics[width=\columnwidth]{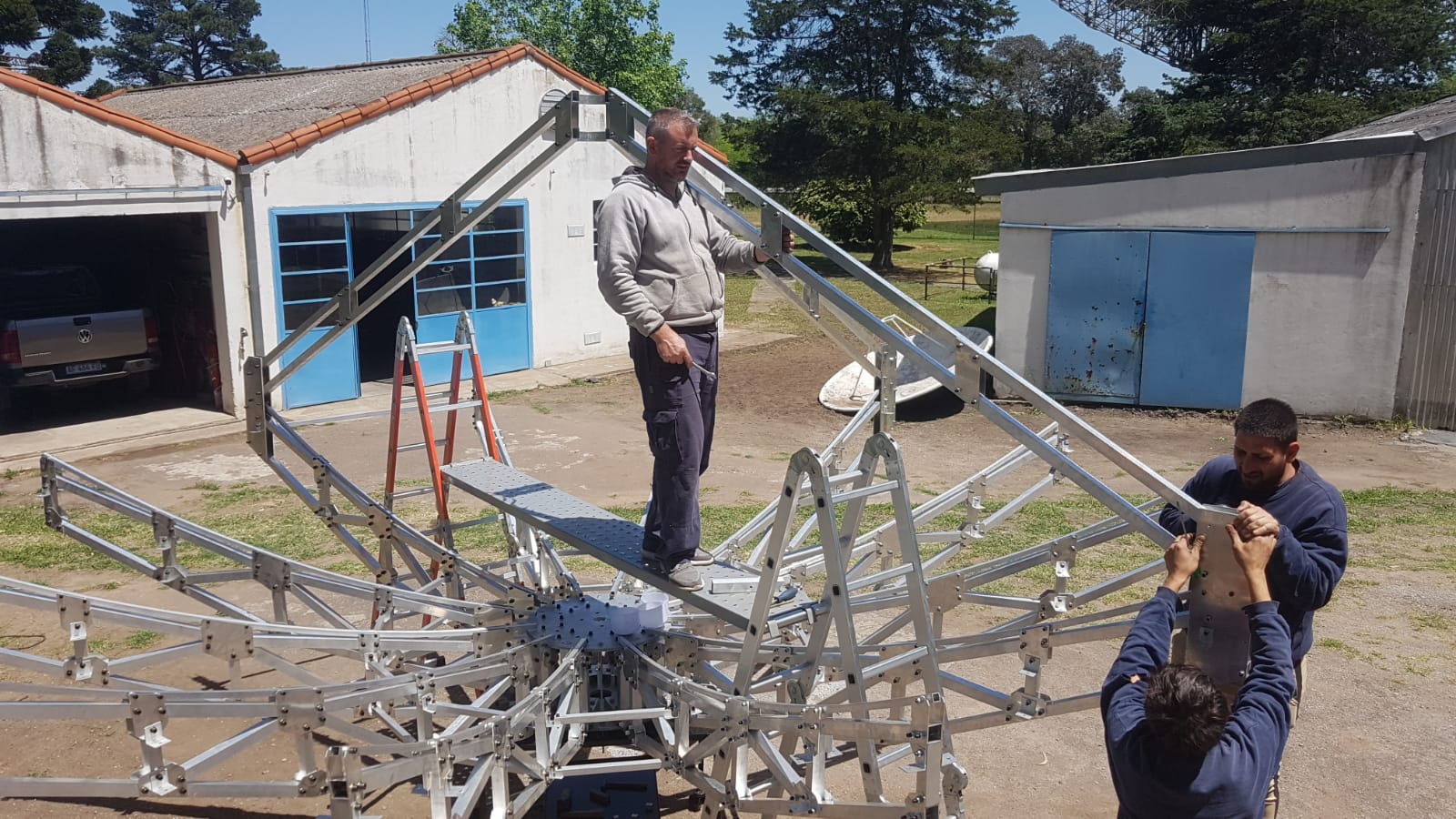}
  \caption{First parabolic dish under construction.}
  \label{fig:ant2}
\end{figure}

\begin{figure}[!t]
  \includegraphics[width=\columnwidth]{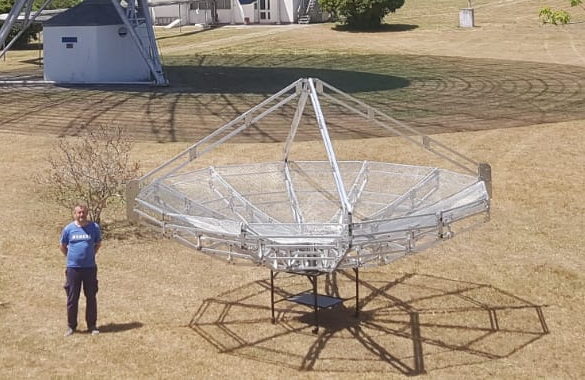}
  \caption{View of the complete parabolic dish.}
  \label{fig:complete}
\end{figure}

The MIA project has reached a major milestone with the construction of the first 5-m parabolic dish. This task represents a significant step forward in the development of the Pathfinder. The technical staff has used its expertise to design the antenna structure and perform simulations using CAD models, as shown in Fig.~\ref{fig:antcad}. The results of these simulations were used to guide the construction phase, as shown in Fig.~\ref{fig:ant2} and Fig.~\ref{fig:complete}. In addition, the mount and support structure for the dish are currently under development. An alt-azimuth mount has been incorporated into the dish design to allow precise pointing and tracking of astronomical sources. The project is focused on developing and perfecting the control systems for the dish to ensure optimal performance.


The digitizer unit is an essential component of the MIA system and the Pathfinder, as it plays a critical role in the acquisition and processing of astronomical data. The primary function of this unit is to acquire the analog signals from the antennas and convert them to digital signals that can be processed by the correlator. To achieve this, the design is based on a SNAP CASPER board, which is a powerful and flexible digital signal processing platform capable of handling large data streams at high speeds.

The SNAP CASPER board consists of a Field Programmable Gate Array (FPGA) programmed to perform real-time digital signal processing on incoming data. The board can be programmed using the CASPER toolflow, which provides a high-level programming language for configuring the FPGA. This language allows custom signal processing algorithms to be designed and implemented to meet the specific requirements of the MIA system.

As mentioned before, the digitizer unit is designed to process a bandwidth of 250 MHz x 3 antennas x 2 polarizations for a total bandwidth of 1500 MHz. The data captured by the digitizer is transmitted over a 10 Gbps network to the correlator where it is correlated to produce astronomical images. The correlator is responsible for combining the signals from each antenna and producing a single output stream representing the astronomical image.

Part of the design is to optimize the performance of the digitizer unit by fine-tuning the CASPER firmware and optimizing the FPGA resources to reduce power consumption and improve overall processing speed. In addition, there is a line of work to develop software tools that can analyze the data acquired by the digitizer unit to identify and mitigate any noise or interference in the system. Overall, the digitizer unit is a critical component of the MIA system.

\section{Summary}
\label{sec:Summary}
We have started the development of a new instrument for radio astronomy research with the aim of expanding the actual scientific capabilities of the IAR.
We are currently studying antenna distribution and optimal array configuration, and in the process of verifying the first electronic modules and system parts of the Pathfinder, such as the front-end, back-end, digitizers, etc. The enterprise will permit us in principle the following:
\begin{itemize}
\item To acquire new know-how on radio interferometry from a technical point of view to add it to our current experience on single dish astronomy; 
\item To increase the know-how and the possibilities for technology transfer;
\item To build a competitive instrument for Argentina and the Latin American scientific community.
\end{itemize}			
In conclusion, the MIA project is a unique and innovative initiative aimed at promoting scientific and technological research in Argentina. The project has been designed to be low cost and to allow the participation of local companies and institutions, which will contribute to the local economy and create opportunities for knowledge exchange. With its high angular resolution and advanced technical capabilities, MIA is expected to make significant contributions to various fields of astrophysics and advance our understanding of the universe.

\section*{Acknowledgments}

We extend our gratitude to the entire staff of IAR for their effort and dedication towards the successful execution of the project objectives. We acknowledge the crucial financial backing provided by the CONICET grants of the program PUE 2020 assigned to the IAR, which has enabled the realization of the present work.

\end{document}